# Scalable single-photon sources in atomically thin MoS$_2$

**Authors:** *J. Klein$^{1,2*}$, L. Sigl$^{1,2*}$, S. Gyger$^{3}$, K. Barthelmi$^{1,2}$, M. Florian$^{4}$, S. Rey$^{1}$, T. Taniguchi$^{5}$, K. Watanabe$^{5}$, F. Jahnke$^{4}$, C. Kastl$^{1,2}$, V. Zwiller$^{3}$, K. D. Jöns$^{3}$, K. Müller$^{1,2}$, U. Wurstbauer$^{6}$, J. J. Finley$^{1,2}$, A. W. Holleitner$^{1,2}$*

$^1$ Walter Schottky Institut und Physik Department, Technische Universität München, Am Coulombwall 4, 85748 Garching, Germany
$^2$ Munich Center of Quantum Science and Technology (MCQST), Schellingstr. 4, 80799 Munich, Germany
$^3$ KTH Royal Institute of Technology, Department of Applied Physics, Albanova University Centre, Roslagstullsbacken 21, 106 91 Stockholm, Sweden
$^4$ Institut für Theoretische Physik, Universität Bremen, P.O. Box 330 440, 28334, Bremen, Germany
$^5$ National Institute for Materials Science, Tsukuba, Ibaraki 305-0044, Japan
$^6$ Institute of Physics, University of Münster, 48149 Münster, Germany
* These authors contributed equally

**Abstract:** Real-world quantum applications, eg. on-chip quantum networks and quantum cryptography, necessitate large scale integrated single-photon sources with nanoscale footprint for modern information technology.[1–3] While on-demand and high fidelity implantation of atomic scale single-photon sources in conventional 3D materials suffer from uncertainties due to the crystals' dimensionality,[4–6] layered 2D materials can host point-like centers with inherent confinement to a sub-nm plane.[7–11] However, previous attempts to truly deterministically control spatial position and spectral homogeneity while maintaining the 2D character have not been realized.[12–14] Here, we demonstrate the on-demand creation and precise positioning of single-photon sources in atomically thin MoS$_2$ with very narrow ensemble broadening and near-unity fabrication yield. Focused ion beam irradiation creates 100s to 1000s of mono-typical atomistic defects with anti-bunched emission lines with sub-10nm lateral and 0.7 nm axial positioning accuracy. Our results firmly establish 2D materials as a scalable platform for single-photon emitters with unprecedented control of position as well as photophysical properties owing to the all-interfacial nature.

Developing scalable approaches for the generation of single-photon emitters (SPEs) in solid-state materials is a longstanding goal in the field of quantum nanophotonics.[4–6,12–16] SPEs attract significant attention due to their prospective use in miniaturized quantum photonic systems. A prominent example is diamond, with the most actively studied color centers being NV and SiV.[2,3] Such color centers are typically induced by ion implantation and subsequent annealing with the SPEs being buried in a bulk matrix several tens of nm below the surface.[4,6] Therefore, their creation is often probabilistic and accompanied by significant lateral and axial positioning inaccuracies. Ultimately, such inaccuracies render their efficient interfacing to nanophotonic circuits challenging and may also deteriorate the emitters' electronic and optical properties, which are impacted by their distance from the surface.[5,6]

A natural workaround is to reduce the physical dimensionality of the material. In this context 2D materials are ideal candidates for hosting SPEs that are inherently restricted to a sub-nm thin material.[17] In recent years, spatially localized luminescence from SPEs in hBN[18], $WSe_2$,[7–11] and $GaSe$[19] has been observed. While deterministic placement of emission centers has been shown in $WSe_2$, $WS_2$ and hBN by creating local exciton confinement potentials,[12–14,20] their 2D character is lost because of the necessary strain profile and the precise origin of emission remains subject to debate. Similarly, localized, yet randomly occurring, single-photon emission has been observed near wrinkles and edges of hBN and GaSe.[15,19] In layered hBN, various attempts to deterministically localize SPEs have been unsuccessful so far owing to the rigidity of the material.[15,21] Moreover, emitters in exfoliated and solution prepared hBN exhibit strong inhomogeneous ensemble broadening and the spatial occurrence of SPEs remains random[16,18,21] most likely reflecting a diverse typology of different defects.[15,18]

Here, we systematically address all current drawbacks of solid-state SPEs that hamper their scalable devices integration by demonstrating deterministic and highly local creation of SPEs in atomically thin $MoS_2$ using a focused He-ion beam. The emitters show narrow optical signatures and significantly reduced ensemble broadening. For generating defects, we utilize a helium ion microscope (HIM) which provides a nanometer-focused stream of He-ions that penetrates through the encapsulation layer without significant loss of its beam properties (see Supplementary Information S1).[22,23] We investigate two types of samples: (i) $MoS_2$ on hBN that is first irradiated *prior to* encapsulation with hBN (Fig. 1a), and (ii) $hBN/MoS_2/hBN$ that is irradiated *after* full hBN encapsulation (Fig. 1b). The encapsulation procedure reduces the background emission from adsorbate-related luminescence (ARL) (Figs. 1c and 1d)[24,25] the so-called *L*-peak emission. Typically after the irradiation of a non-encapsulated monolayer $MoS_2$, accumulation of adsorbates on the defective surface like $N_2$, $O_2$ and $H_2O$ is expected,[22] that can manifest in chemically altered defect state morphology or simply through ambient passivation of sulfur vacancies with oxygen.[26,27] Figure 1c demonstrates that for local implantation through the readily assembled hBN/MoS$_2$/hBN heterostructure the luminescence features the emission from excitons (X) and a single SPE with virtually no background emission. The high optical quality of our emitters is a consequence of the reduction of contaminants from the van der Waals interfaces that are squeezed out during the stacking process resulting in electronically homogeneous heterointerfaces.[24]

We investigate the emission statistics of SPEs obtained from spatially resolved µ-PL mappings of nominally identical samples for irradiation *prior to* and *after* hBN encapsulation. For both sample configurations, we irradiate circular areas with diameters ranging from 700 nm down to 100 nm with a pitch of 2 µm and a constant He-ion dose of $\sigma = 5 \cdot 10^{12}$ ions cm$^{-2}$, which is far below the threshold for modifying the free exciton emission.[22] By reducing the irradiated area, we demonstrate emission from few SPEs down to a single SPE within the resolution of our laser spot ($d_{Laser}$ ~ 700 nm). Figure 2a shows spatially resolved µ-PL mappings of a hBN/MoS$_2$/hBN of type (i) for exposure diameters of 700 nm, 400 nm and 200 nm. The PL is spectrally integrated between 1.7 eV and 1.88 eV to reveal emission from He-ion induced SPEs (Supplementary Information SI Fig. 2). The He-ion irradiation manifests in spatially activated SPEs clearly following the imprinted writing pattern. Typical spectra of fields with 700 nm diameter reveal multiple overlapping SPEs that are within the point spread function of our laser spot. However, for the smaller diameters ≤ 200 nm, we observe distinct individual emission lines at ~ 1.75 eV corresponding to an energy detuning of $\Delta E = E(X) - E(SPE) = (194 \pm 1)$ meV. These lines correspond to individual single photon emitters,[23] as will be shown below. Importantly, we consistently find the ARL to be reduced for He-ion irradiation through the fully assembled heterostructures. For comparison, Fig. 2c shows a µ-PL mapping of a nominally identical sample but of type (ii). Here, the quality of the spectra is limited by the spectrally broad ARL, and the SPEs are distributed more widely in emission energy as evident from the emission statistics obtained from peak finder algorithms. Statistics obtained from the sample irradiated *after* encapsulation (type (i) in Fig. 2e) reveals a narrow inhomogeneous ensemble of SPEs at an energy detuning $\Delta E = (194 \pm 1)$ meV and with a FWHM of $(28 \pm 1)$ meV. This is in strong contrast to samples of type (ii) (Fig. 2f). Here, the emission distribution at the main energy detuning is significantly broadened with a FWHM of $(40 \pm 4)$ meV, and a second, similarly intense emission distribution at $\Delta E = (139 \pm 2)$ meV emerges with a FWHM of $(52 \pm 5)$ meV. This second emission is almost entirely absent in the irradiated sample *after* encapsulation (Fig. 2e), and we attribute it to stem from randomized interactions of the SPEs with molecular species during the ambient exposure (Fig. 1b).[25]

We continue to determine the probability and yield to activate single-photon emission in the He-ion irradiated circular areas. We expect the number of emission centers and their corresponding intensity in first approximation to scale linearly with the number of ions, and hence also with the irradiated area at a constant area dose. Indeed, the mean integrated SPE intensity exhibits a linear dependence on the corresponding irradiated area, or equivalently absolute number of ions for small exposure areas (Fig. 2g). Moreover, statistically evaluating the percentage of activated fields (Fig. 2h), we find unity creation efficiency for 700 nm for $MoS_2$ irradiated *prior to* encapsulation, and the activation efficiency for irradiated $MoS_2$ *after* encapsulation reaches values > 70 %. This discrepancy is very likely due to a substantial number of emitters at an energy detuning of $\Delta E =$ 139 meV (Fig. 2f) that are absent in irradiated $MoS_2$ *after* encapsulation. The origin of such emission may be either due to 'dark' He-ion induced defects that are activated through interaction with functional groups or modified 'bright' emitters at the detuning $\Delta E =$ 194 meV, where a complex wave function admixture shifts the emission energy. The latter possibility suggests interesting prospects for intentional manipulation of SPEs in 2D materials by their interaction with external species.

The SPEs implanted in the fully encapsulated $MoS_2$ show superb optical properties with narrow zero phonon lines (Fig. 3a) at 10 K and a weak satellite peak at a Raman shift of ~ 248 cm$^{-1}$ (energy detuning of ~ 31 meV). Intriguingly, the latter line does not match with optical phonon energies,[28] but it may likely be explained by a local phonon mode (LPM) of the defect (Supplementary S3). To prove the non-classical nature of light emission from our He-ion implanted defects, we perform a Hanburry Brown and Twiss (HBT) experiment. Figure 3b depicts the second-order intensity autocorrelation function $g^{(2)}(\tau)$ of an implanted defect under CW excitation measured at 5 K. We clearly observe anti-bunching with a $g^{(2)}(0) = 0.23 \pm 0.04$ at zero time delay ($\tau = 0$). Moreover, the fit suggests a long radiative recombination time of $\tau = (1.73 \pm 0.15)$ µs, which contrasts strongly with the ns lifetimes typically observed for other localized individual emitters in all other 2D materials[7–11,18,19] but is similar to cold atoms and trapped ions. At lower bath temperatures (Fig. 3c), the zero phonon line narrows significantly to 248 µeV at $T = 1$ K. The data are well described with an independent boson model fit[23] (inset Fig. 3c) resulting in a Debye-Waller factor of 0.27. The excellent agreement with this model further suggests that the SPE in $MoS_2$ can be well understood with atomic physics, similar to color centers in diamond. Moreover, the fit suggests a homogeneous (inhomogeneous) contribution of 100 µeV (188 µeV) to the linewidth. This is still larger than the transform limit, which can be explained by the non-resonant excitation, as for solid-state SPEs transform-limited linewidth is typically only observed under resonant excitation.

To estimate the lateral and axial precision with which we generate SPEs in the $MoS_2$ monolayer, we perform Monte-Carlo simulations of beam trajectories through a thick slab of hBN to obtain the beam resolution of the primary beam (Supplementary Information SI Fig. 1). For typical hBN thicknesses used throughout this manuscript (< 20 nm), we obtain a primary He-ion beam resolution of < 5 nm at the monolayer. Recent numerical calculations proposed that the lateral resolution to create defects in a substrate-supported monolayer $MoS_2$ for the beam energies (30 keV) used in our samples is < 10 nm.[29] Overall, we anticipate that secondary ion processes will be the limiting factor in $MoS_2$ layers supported on thick substrates.

We implant SPEs in a readily assembled van der Waals heterostructure using spot exposures, where the exposed area only corresponds to the area of the ion beam (Figure 4). To obtain reliable statistics, we wrote arrays of 4 × 10 exposures with a pitch of 2 µm (as highlighted for field number six from the top in Fig. 4a). Figure 4a shows a spectrally integrated PL map of such an irradiated sample. The number of ions per spot is varied from 400 to $329 \cdot 10^3$ (top to bottom) for the different arrays. The monolayer $MoS_2$ is overlaid in purple for clarity. In this way, we identify the optimal dosage and SPE yield. Moreover, the letters 'TUM' are written with a constant dose of $\sigma = 0.3 \cdot 10^{13}$ cm$^{-2}$ and are clearly visible as bright emission from an ensemble of SPEs. By analyzing each irradiated position, we can statistically evaluate the overall percentage of activated fields for each set of doses and the number of SPEs per spot (Fig. 4b-g). The probability distribution for selected doses ranging from 400 to $329 \cdot 10^3$ ions per spot and the corresponding irradiated areas are highlighted in Fig. 4a by the same colors. The total yield of activated fields (Fig. 4h, black) increases monotonically from 20 % to 60 % throughout the range of doses used for He-ion irradiation. From the activated fields, we find that for the lowest dose of 400 ions per single irradiation (Fig. 4b), about 80 % of spots showing luminescence manifest in the generation of one SPE. For larger doses, this number decreases to 37 % for the highest number of ions (Fig. 4h, cyan). Typical spectra for selected doses are shown in Fig. 4g. All spectra show a SPE at a very similar detuning from the neutral exciton X. For lower doses of < 5100 ions per spot, obtaining a single SPE has the highest relative probability (~ 80 %), combined with narrow inhomogeneous ensemble broadening due to implantation through the fully encapsulated heterostructure.

The ability for single-photon source creation in already assembled van der Waals devices opens up a manifold of opportunities to investigate and control their optical properties in a context that is fully compatible with top down nanofabrication (Supplementary Table S4). In particular, implantation in as-fabricated field-switchable van der Waals devices for potential control of the defects charge state, integration in cavities and waveguide with high Purcell enhancements for potential room-temperature operation or scalable positioning for investigating superradiance effects of coherently coupled single-photon sources are straightforwardly envisioned for the near future.

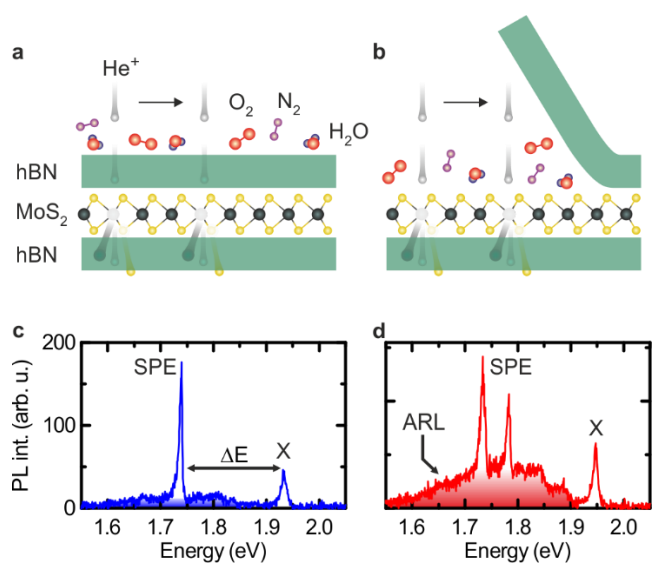

**Fig. 1. | Atomically thin MoS$_2$ as a platform for scalable and background free single-photon emitters. a,** Impact of layer sequence of the investigated van der Waals heterostructures and focused He-ion SPE generation. He-ion irradiation *after* hBN encapsulation (type i) results in clean interfaces with significantly reduced contamination from adsorbates. **b,** He-ion irradiation *prior to* encapsulation (type ii) promotes the interaction of defects with ambient contaminants. **c,** Typical low temperature (15 K) PL spectrum of irradiated hBN/MoS$_2$/hBN *after* encapsulation. The spectrum features neutral exciton emission (X) and the emission of a single-photon emitter (SPE). Adsorbate-related luminescence (ARL) is suppressed owing to the hBN encapsulation. **d,** The PL spectrum from a heterostructure exposed *prior to* encapsulation reveals inhomogeneous single-photon emission with multiple emission lines distributed over a broader spectral window and significant ARL.

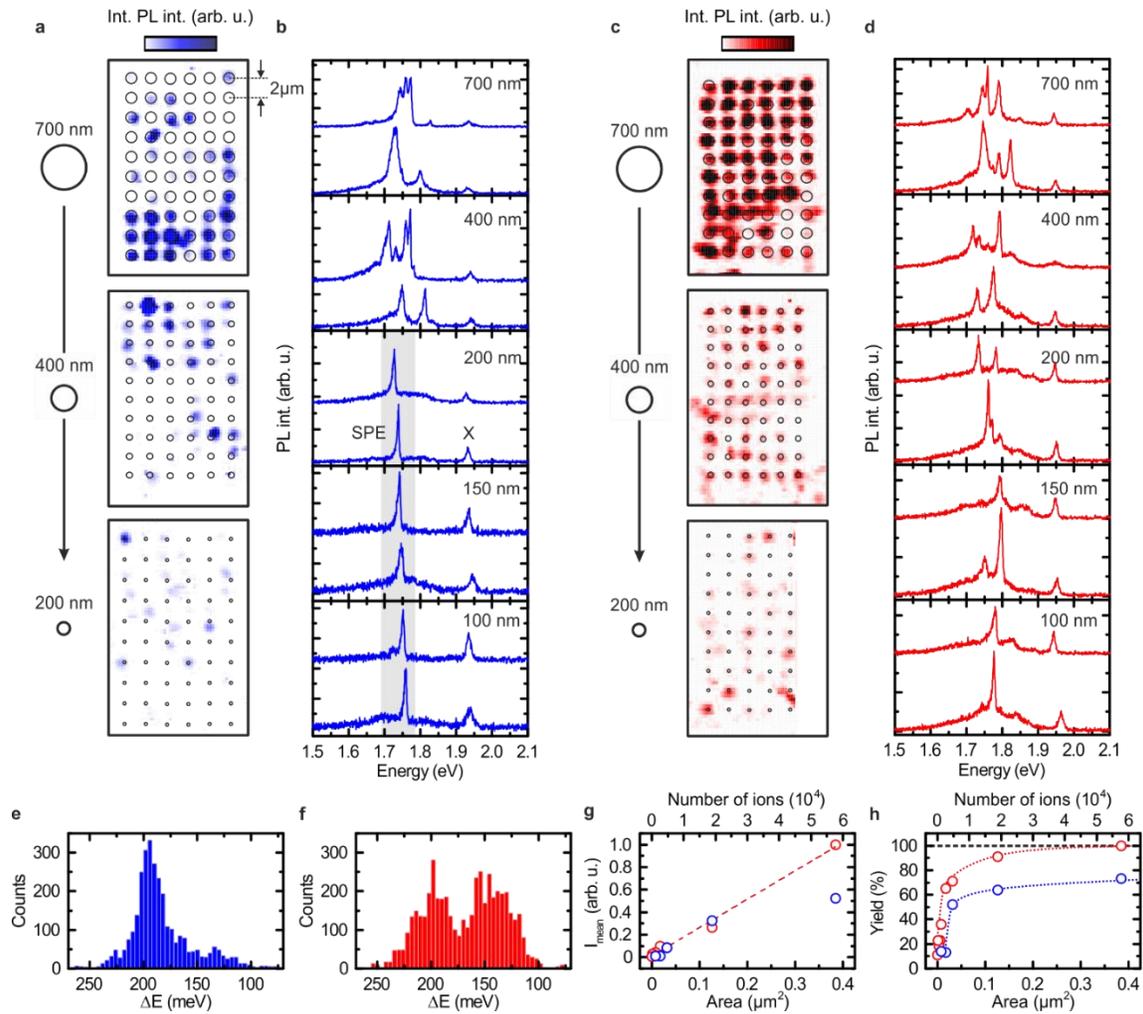

**Fig. 2 | Ensemble generation of spectrally clean single-photon emitters with unity yield and narrow inhomogeneous ensemble broadening. a,** Spatially resolved low temperature (15 K) µ-PL spectroscopy of SPEs in locally He-ion irradiated hBN/MoS$_2$/hBN (type i). The ion dose is kept constant at σ = 5 · 10$^{12}$ ions cm$^{-2}$ and the exposed area is varied. Circular exposures with varying diameter are arranged in a matrix with a pitch of 2 µm. **b,** Typical PL spectra for circular areas of varying diameter are shown (700 nm, 400 nm, 200 nm, 150 nm and 100 nm). Spectra from the largest areas exposed contain multiple SPEs, whereas fields with d ≤ 200 nm are in the limit of an individual SPE with reproducible spectral emission energies (~ 1.75 eV) and suppressed ARL. **c,** µ-PL mapping of SPEs in locally He-ion irradiated MoS$_2$ *prior to* encapsulation (type ii). The same dose and diameters were used. **d,** Typical PL spectra for circular areas of varying diameter are shown. Spectra exhibit multiple SPEs that are widely distributed and energy with significant ARL. **e,f,** Inhomogeneous ensemble broadening (energy detuning *ΔE* = *E*(X)-*E*(SPE)) of emission. Irradiated MoS$_2$ *after* encapsulation exhibits a narrow inhomogeneous ensemble broadening (FWHM = (28 ± 1) meV centered at a detuning of *ΔE* = (194 ± 1) meV. Irradiation *prior to* full encapsulation reveals a larger ensemble broadening (FWHM = (40 ± 4) meV) with an additional spectrally broad distribution (FWHM = (52 ± 5) meV) centered at *ΔE* = (139 ± 2) meV. **g,** Mean integrated SPE intensity per activated field. The data suggests a linear relation of intensity and area of irradiated MoS$_2$ *prior to* encapsulation. The deviation of the last data point for the irradiated fully encapsulated MoS$_2$ is attributed to sample inhomogeneities. **h,** Creation yield of He-ion irradiated fields as a function of area. Unity yield is highlighted with the dashed horizontal line. The colored lines are a guide to the eye.

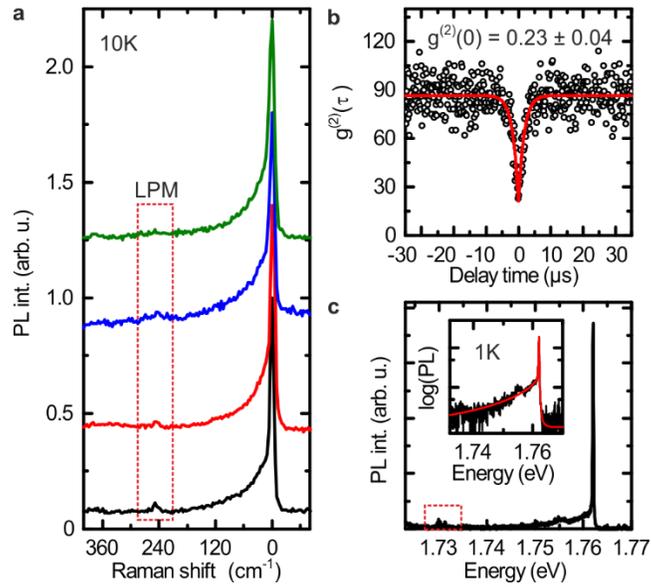

**Fig. 3 | Photon anti-bunching of He-ion implanted single-photon sources in atomically thin MoS$_2$. a,** Photoluminescence spectrum of four typical SPE at 10 K with a FWHM of ~ 1.42 meV. The local phonon mode (LPM) at a Raman shift of ~ 248 cm$^{-1}$ (energy detuning of ~ 31 meV) of the SPE is highlighted. **b,** Second order correlation $g^{(2)}(\tau)$ function of a SPE showing clear anti-bunching with $g^{(2)}(0) = 0.23 \pm 0.04$ and a lifetime of $\tau = 1.73 \pm 0.15$ µs at 5 K. **c,** Photoluminescence spectrum of a SPE at 1 K with a FWHM of 248 µeV and a LPM at around ~ 248 cm$^{-1}$. Inset: Same data on logarithmic scale fitted with an independent boson model.

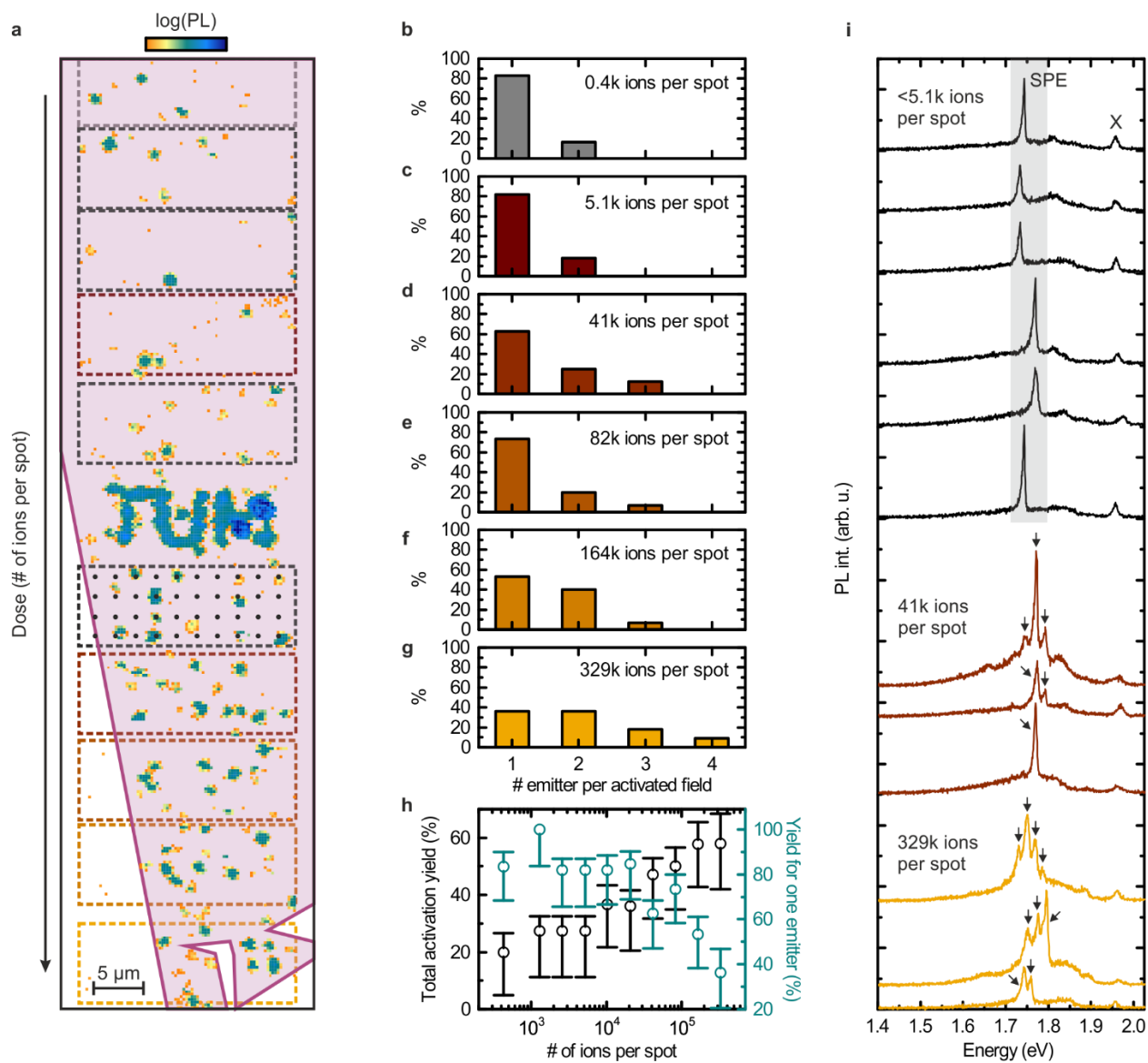

**Fig. 4 | Deterministic generation of single-photon sources with sub-10nm lateral resolution.**
**a,** Spatially resolved and spectrally integrated low temperature (15 K) µ-PL mapping of a monolayer $MoS_2$ (purple area). The PL intensity is integrated in the spectral window 1.7 to 1.88 eV to reveal He-ion induced single-photon emission. Defects are implanted through the hBN/$MoS_2$/hBN heterostructure with a tightly focused He-ion beam. The thickness of the top hBN layer is < 15 nm. Arrays of He-ion point shots are arranged with a pitch of 2 µm (black dots) while the dose (number of ions per spot) increases from top to bottom. The 'TUM' letters are written with a constant dose of $\sigma = 0.3 \cdot 10^{13}$ ions cm$^{-2}$ to obtain ensemble defect emission. **b-g,** Probability distribution of the number of SPEs per irradiated spot for different dosages ranging from 400 to $329 \cdot 10^3$ ions per spot. For the lowest number of ions, about 80 % of the activated spectra exhibit a single emission line, while fields exposed to higher numbers of He-ions are more likely to show multiple defect emission lines. **h,** Total percentage of activated fields (black) and likelihood to obtain a single emitter within activated spots (cyan). **i,** Typical PL spectra from **a** reveal the neutral exciton X at 1.95 eV and SPE emission. For low doses (< 5100 ions), spots show typically single-photon emission ~ 1.75 (highlighted in gray) while fields with higher dose ($41 \cdot 10^3$ and $329 \cdot 10^3$ ions per spot) are more likely to show multiple emitters as highlighted by arrows.

**Correspondence and requests for materials** should be addressed to:

julian.klein@wsi.tum.de, finley@wsi.tum.de or holleitner@wsi.tum.de

**Data and materials availability:** The data presented in this manuscript are available from the authors upon reasonable request

**Competing interests:** The authors declare no competing interests

**Methods**

**Optical spectroscopy**

For the spatially resolved PL data, a continuous-wave 488 nm laser was used for excitation. The sample was cooled down to 10K in a continuous flow cryostat (Janis ST-500). The laser was focused onto the sample with a microscope objective (NA = 0.42, 50x, Mitutoyo). Scanning was performed by a two axis scanning galvanometer mirror system (Thorlabs) combined with a 4f lens geometry. The emitted light was separated from the excitation path by a 50:50 beamsplitter and dispersed in a spectrometer (Spectra Pro HRS-300) and projected onto a charge coupled device CCD (PIXIS). The excitation laser was blocked by a longpass filter (RazorEdge LP 488).

The sample at 5K (attoDry800) was excited using a HeNe laser. The emission was collected with a (50x, NA = 0.81, attocube) microscope objective, fiber-coupled and filtered using two tunable band-pass filter. The fiber-based HBT setup consists of a 50:50 beamsplitter and two superconducting-nanowire single-photon detectors (Single Quantum) with efficiencies of 50 %, 60 %, timing jitters of 20 and 30ps, and dark count rates of 0.006 and 0.017 cts/s.

µ-PL - Spectroscopy at 1K was performed in a cryogen-free dilution refrigerator operated at an elevated temperature of 1K. Optical side access through anti-reflection coated windows was used to focus a continuous-wave 632nm laser on the sample using a (50x, NA = 0.82, Partec) microscope objective.

**Author Contributions**

J.K. and L.S. contributed equally.

J.K., L.S., C.K., K.M., U.W., J.J.F. and A.W.H. conceived and designed the experiments, T.T. and K.W. provided hBN crystals, S.R. prepared the samples, J.K. performed He-ion irradiation, L.S., J.K., S.G., K.D.J. and K.B. performed the optical measurements, L.S., J.K., S.G., V.Z. and K.D.J. analyzed the data, M.F. and F.J. modelled spectral lineshapes, J.K. wrote the manuscript with input from all authors. All authors reviewed the manuscript.

**Acknowledgements**

The work was supported by Deutsche Forschungsgemeinschaft (DFG) through the German Excellence Strategy via the Munich Center for Quantum Science and Technology (MCQST) - EXC-2111–390814868 and e-conversion – EXC 2089/1 – 390776260. We gratefully acknowledge support through TUM International Graduate School of Science and Engineering (IGSSE). We also gratefully acknowledge financial support from the European Union's Horizon 2020 research and innovation program under grant agreement No. 820423 (S2QUIP) the German Federal Ministry of Education and Research via the funding program Photonics Research Germany (contract number 13N14846) and the Bavarian Academy of Sciences and Humanities. J.K. acknowledges support from the Alexander von Humboldt Foundation. S.G. acknowledges funding from the Swedish Research Council under grant agreement 2016-06122 (Optical Quantum Sensing). K.D.J. acknowledges funding from the Swedish Research Council (VR) via the starting Grant HyQRep (Ref 2018-04812) and The Göran Gustafsson Foundation (SweTeQ). M.F. and F.J. acknowledges support from the Deutsche Forschungsgemeinschaft (RTG 2247 "Quantum Mechanical Materials Modelling"). K.W. and T.T. acknowledge support from the Elemental Strategy Initiative conducted by the MEXT, Japan and and the CREST (JPMJCR15F3), JST.